\documentclass[11pt,a4paper]{article}
\usepackage{amsmath,amssymb,amsthm}
\usepackage{epsfig}
\usepackage{psfrag}
\usepackage{diagrams}

\usepackage{chicago}          
\renewenvironment{proof}[1][\proofname]{\par \normalfont \trivlist
\item[\hskip\labelsep\itshape #1]\ignorespaces
}{%
\hspace*{\fill}$\Box$ \endtrivlist }
\renewcommand{\proofname}{{\bf Proof}}

\newcommand{\R}{\mathbb{R}}
\newcommand{\RR}{\mathbb{R}}
\newcommand{\PP}{\mathbb{P}}
\newcommand{\Q}{\mathbb{Q}}
\newcommand{\E}{\mbox{\textnormal{E}}}

\newcommand{\EQ}{\E_\Q}
\newcommand{\half}{\frac{1}{2}}

\newtheorem{prop}{Proposition}

\newcommand{\captionfonts}{\small}

\makeatletter  
\long\def\@makecaption#1#2{%
  \vskip\abovecaptionskip
  \sbox\@tempboxa{{\captionfonts #1: #2}}%
  \ifdim \wd\@tempboxa >\hsize
    {\captionfonts #1: #2\par}
  \else
    \hbox to\hsize{\hfil\box\@tempboxa\hfil}%
  \fi
  \vskip\belowcaptionskip}
\makeatother   

\title{Multivariate Feller conditions in term structure models: Why do(n't) we care?\thanks{The views in this paper are those of the individual authors and do not necessarily reflect official
positions of De Nederlandsche Bank.}}
\author{Peter Spreij\thanks{Korteweg-de Vries Institute for Mathematics, Plantage Muidergracht 24, 1018TV Amsterdam, The Netherlands, Email: spreij@science.uva.nl}
\and Enno Veerman\thanks{Korteweg-de Vries Institute for Mathematics, Plantage Muidergracht 24, 1018TV Amsterdam, The Netherlands, Email: eveerman@science.uva.nl}
\and Peter Vlaar\thanks{Financial Research Department, De Nederlandsche Bank, P.O.Box 98, 1000AB Amsterdam, The Netherlands, Email: p.j.g.vlaar@dnb.nl.} }

\begin{document}

 \maketitle \thispagestyle{empty}
\begin{abstract}
In this paper, the relevance of the Feller conditions in discrete
time macro-finance term structure models is investigated. The
Feller conditions are usually imposed on a continuous time
multivariate square root process to ensure that the roots have
nonnegative arguments. For a discrete time approximate model, the
Feller conditions do not give this guarantee. Moreover, in a
macro-finance context the restrictions imposed might be
economically unappealing. At the same time, it has also been
observed that even without the Feller conditions imposed, for a
practically relevant term structure model, negative arguments
rarely occur. Using models estimated on German data, we compare
the yields implied by (approximate) analytic exponentially affine
expressions to those obtained through Monte Carlo simulations of
very high numbers of sample paths. It turns out that the
differences are rarely statistically significant, whether the
Feller conditions are imposed or not. Moreover, economically the
differences are negligible, as they are always below one basis
point.
\end{abstract}

\clearpage

\section{Introduction}

This paper addresses the necessity of the so-called Feller
conditions for an affine multivariate term structure model with
time-varying volatility. Although already interesting from a
theoretical point of view, the main motivation behind this study
stems from sometimes unattractive implications of the Feller
conditions in macro-finance models. A term structure model in
general involves one or more driving factors, which are usually
assumed unobservable. In the macro-finance models, the driving
factors involve macro-economic variables, for instance the
inflation rate. As these driving factors have a direct economic
interpretation, the restrictions implied by the Feller conditions
can be unappealing from an economic point of view. Consequently,
most macro-finance term structure models assume a constant
volatility for the driving factors, see for instance \citeN{cv02},
\citeN{ap03}, \citeN{ab04}, Dewachter, Lyrio and Maes (2004,
2006)\nocite{dlm04}\nocite{dlm06}, \citeN{fen05}, \citeN{brs05},
\citeN{dl06}, \citeN{htv06}, \citeN{wu06}, and
\citeN{rw07}.\footnote{An exception is \citeN{spe04}, who
specifies a 10-factor model for the US yield curve, including one
heteroscedasticity factor which is a linear combination of several
macroeconomic variables.} This however implies that interest rates
are assumed symmetric, which means that either very low interest
rates are predicted too often or very high interest rates not
often enough. Especially for asset liability management purposes
for pension funds, these characteristics can easily lead to wrong
conclusions as the long duration of their liabilities makes them
extremely sensitive to interest rate changes.

The Feller conditions serve several purposes in a continuous time
affine term structure model with time-varying volatility. First of
all, the dynamics of the factor processes involve square root
terms and the conditions are sufficient to have the arguments of
the square roots strictly positive. This in turn guarantees that
the stochastic differential equations that describe the dynamics
of the factor have a unique strong solution and that a closed form
expression for the bond price can be obtained.

In practice, one often works with discrete time models, which, for
instance, can be obtained by discretizing a continuous time model.
 For
a discrete time model, existence of (strong) solutions is clearly
not an issue and the Feller conditions play no role in this
context. Moreover, the Feller conditions applied to a discretized
model are useless to guarantee that the square roots always have
nonnegative arguments. Indeed,  the standard normally distributed
errors, that are used as inputs in these discrete time models,
imply that at each time instant there is a positive probability
that one or more of the arguments of the square roots become
negative, regardless whether the Feller conditions are satisfied
or not.\footnote{This has already been observed
in~\citeN[p.~290]{bft01} (one of the first papers with an affine
model in discrete time) for a one-dimensional process, although
curiously enough, in the same paper it is claimed that the
multivariate Feller conditions are sufficient for nonnegative
arguments.} \footnote{An alternative would be to assume a Poisson
mixture of Gamma distributions \cite{dls05} for the volatility
factor instead. For a macro-finance model, this is problematic
however, as volatility is an unknown linear function of the
underlying data. Imposing the volatility factor to be equal to one
of the driving factors (for instance expected inflation) has the
disadvantage that this factor is not allowed to be influenced by
the other factors (for instance the real short term interest
rate).} In spite of all this, it is not uncommon to impose the
Feller conditions on discrete time models, as they serve as
approximations of continuous time models.

\begin{figure}\label{fig:nofeller}
\begin{center}
\includegraphics[width=10cm]{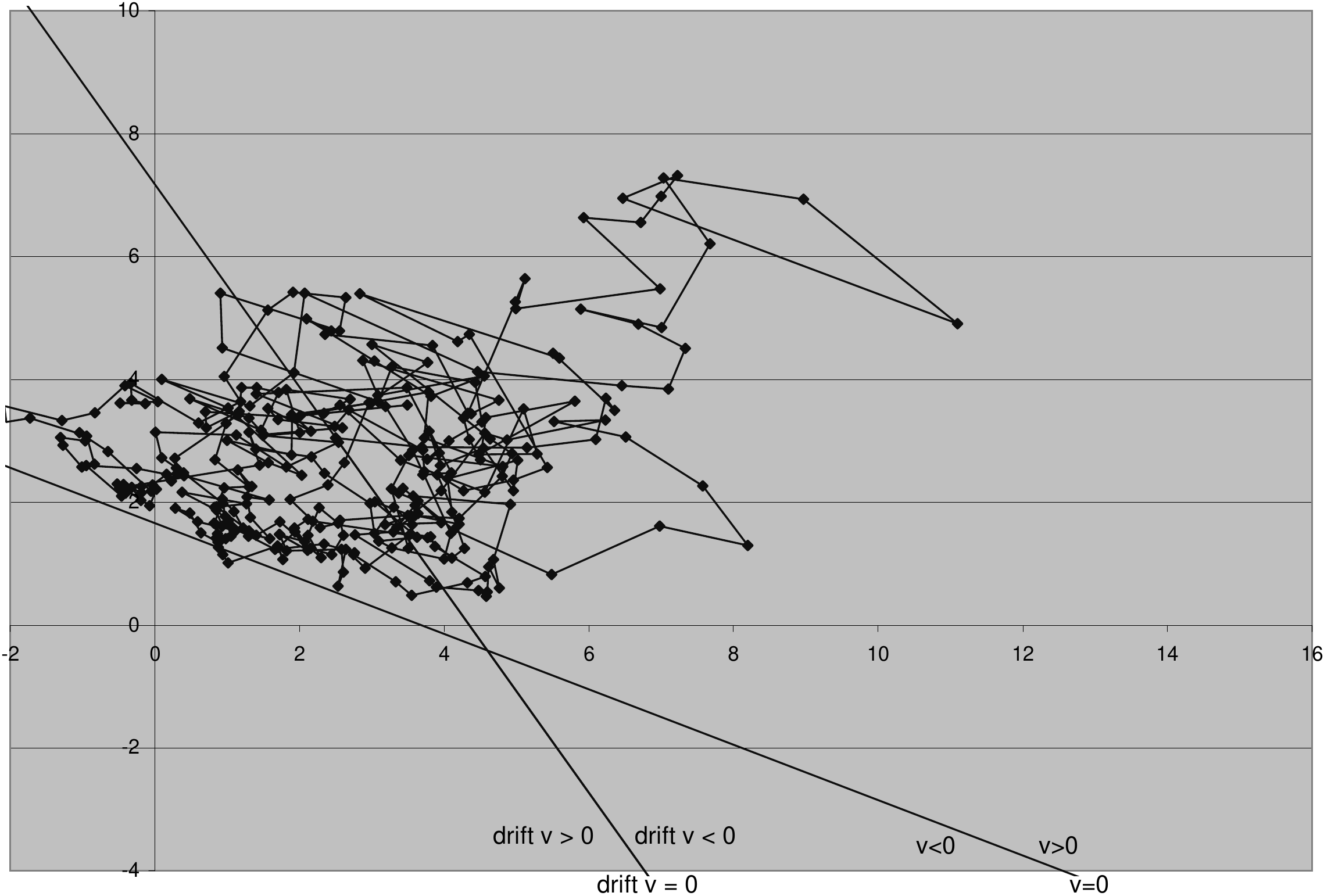}
\end{center}
\caption{Example of a sample path of a discrete time ATSM for
which the Feller conditions are not satisfied.}
\end{figure}

In a two-dimensional setting with one volatility factor (denoted
$v$), one of the Feller conditions imposes that for every point on
the line $v=0$, the deterministic part of the process (the drift) is
such that the volatility becomes positive again. Although it is
clear that this condition is necessary in a univariate setting, its
significance in a multivariate setting is not obvious as the
interaction between the factors limits the part of the line $v=0$
that is actually approached. To illustrate this,
Figure~\ref{fig:nofeller} shows a typical trajectory of a two-state
factor process for a discrete time model with one volatility factor,
and where the Feller conditions are not imposed. Although the line
$v=0$ is crossed once, this happens in the area where the drift of
the volatility process is positive, which would be impossible in a
continuous time model. In other words, the fact that $v$ also
assumes negative values is due to the discretization of the model,
not because the Feller conditions are violated.

All this motivates a study that sheds some light on the necessity to
impose the Feller conditions. If one assumes for a continuous time
model that all volatilities are positive (which happens if the
Feller conditions hold), there is an exponentially affine formula
that expresses bond prices in terms of the state factor. The
coefficients in this formula can be expressed in the model
parameters. For a discrete time model, one can algebraically derive
a similar expression, but again under the assumption that the square
root factors are always positive. Since the latter does not hold
(even when the Feller conditions are imposed),  an exponential
affine formula in discrete time has to be considered as yet another
approximation for the bond price rather than an exact expression.

In order to investigate the accuracy of such an approximation, we
compare them with an alternative method for validating bond prices,
by making use of Monte Carlo simulations. This will be done for a
multi-factor discrete time term structure model with expected
inflation and the real short rate as factors. Such a model will be
estimated using European data.

Since expected inflation is not observed, we use the Kalman filter
combined with a likelihood approach to estimate the involved
parameters. Estimation is performed for a number of cases, that
will be referred to as models with {\em independent volatilities},
{\em dependent volatilities} and {\em proportional volatilities}.
Each of these models will be estimated, with and without the
Feller conditions imposed. In a pure latent variable model, it is
usual to impose  these conditions by assuming a \emph{canonical
form}, as in \citeN{ds00}. In such a canonical form the volatility
factors are equal to some of the state factors. However, we cannot
do this for a macro-finance model as ours, since none of the
factors can be taken as a volatility factor a priori. Therefore we
will extract explicit parameter restrictions from the Feller
conditions for our non-latent variable model.

Having executed the estimation, we compare two consecutive
approaches to validate bond prices. In the first one, we calculate
the bond prices directly given the estimated parameters, using the
Riccati equations (thereby ignoring the cut-off of volatility at
zero). In the second approach, we perform a high number of Monte
Carlo simulations of the trajectories of the factors, whereby
volatility is restricted to be nonnegative. The mean of these
simulations gives a second approximation of the bond price.
Moreover, we can measure the approximation error with (sampled)
confidence intervals. We will show that the differences between
Monte Carlo results and the values obtained from the exponential
affine formula are almost always negligible, both economically and
statistically, whether
the Feller conditions are imposed or not. From an economic point of view,
the difference in implied yields
between the two methods is hardly relevant, as it is at most one
basis point. Statistically, the difference is only significantly
different from zero for some maturities for the dependent and
independent volatility models without Feller conditions. For the
proportional volatility models, there is never a problem.

The rest of this paper is organized as follows. In
Section~\ref{sec:atsm} we review general and affine term structure
models in continuous time, mainly with the aim to set the notation
for the sections to follow. Moreover, we explicitly pin down the
Feller conditions for a two-dimensional affine model, since the
model is not given in canonical form. In Section~\ref{sec:ctod} we
discretize a continuous time model and show that the discretized
model leads to the same expression for bond prices as a the
discretized version of formula for bond prices in continuous time.
In Section~\ref{sec:model} we present and estimate our models. In
Section~\ref{sec:mc} we use the estimated models of
Section~\ref{sec:model} to price bonds by Monte Carlo methods and
compare the obtained results with those obtained by analytic
methods. Finally in Section~\ref{sec:conclusions} we summarize our
findings and draw
some conclusions.

\section{Affine term structure models in continuous
time}\label{sec:atsm}
Although we propose an affine term structure
model for \emph{discrete} time, we first discuss \emph{continuous}
time models. The reason for this is that the mathematical theory for
affine term structure models was initially developed for continuous
time models \cite{dk96}, whence in this respect it is natural to
regard the equations governing the discrete time model
 as discretizations of the continuous
time equations. In particular the Feller conditions we are
interested in apply to continuous time models only. In order to
understand the necessity of this condition in discrete time, we
first have to understand the reason why it should be imposed for
continuous time and then we can check whether these reasons still
apply after discretization. Furthermore, since the Feller
conditions use continuous time parameters, exact knowledge of the
correspondence between discrete and continuous time parameters is
required in order to impose it on the discrete time model.

\subsection{Short rate term structure models}\label{cont}

Let us first concisely recall some general theory of short rate
term structure models, see \citeN{Hunt}, \citeN{musrut} or
\citeN{brimer} for details.  The formulas below will be used in
subsequent sections. We assume that all relevant expressions are
well defined.

In a short rate term structure model the price $D_{t,T}$ of a
zero-coupon bond at time $t$ maturing at $T$ is based on the
dynamics of the short rate $r$ through the formula
\begin{equation}\label{eq:cbpR}
D_{t,T}=\EQ(\exp(-\int_t^T r_s ds)|\mathcal{F}_t),
\end{equation}
with $\Q$ the risk-neutral measure and $(\mathcal{F}_t)$ the
underlying filtration.

Typically, in a short rate model one chooses $r$ to be a function
of a (possibly multi-dimensional) process $X$ which satisfies a
stochastic differential equation~(SDE)
\[
dX_t = \mu(t,X_t)dt+\sigma(t,X_t)dW^\Q_t,
\]
with $W^\Q$ a multivariate Brownian motion under the risk-neutral
measure $\Q$ and one writes $r_t=r(X_t)$.

Under rather general conditions there exists a strong solution $X$
to this equation which is Markov. In this case the bond-price can
be written as $D_{t,T}=\EQ(\exp(-\int_t^T r(X_s)
ds)|X_t)=:F(t,X_t)$ for some function $F$. If $F$ is smooth
enough, then it solves the {\em fundamental partial
differential equation} (PDE), also called \textit{term structure
equation} (see~\citeN[Chapter 12]{musrut} or~\citeN{vas77}, where
the latter terminology was introduced)
\begin{equation}\label{eq:pde}
\frac{\partial}{\partial
t}F(t,x)+\mathcal{L}F(t,x)-r(x)F(t,x)=0,\quad F(T,x)=1,
\end{equation}
with
\[
\mathcal{L}=\sum_i \mu_i \frac{\partial}{\partial
x_i}+\half\sum_{i,j}(\sigma\sigma^\top)_{ij}\frac{\partial^2}{\partial
x_i\partial x_j},
\]
the generator of $X$, where $\sigma^\top$ means the transpose of
$\sigma$.

The physical measure $\PP$ is equivalent to the risk-neutral
measure $\Q$ and related via a density process $L$ by
$
\left.L_{t}=\frac{d\PP}{d\Q}\right|_{\mathcal{F}_t}.
$
The process $L$ can often be written as an exponential process
$\mathcal{E}(Y\cdot W^\Q)$ for some $Y$, i.e.
\begin{equation}\label{eq:L}
L_t = \exp(\int_0^t Y^\top _s dW^\Q_s - \frac{1}{2}\int_0^t
Y_s^\top Y_s ds),
\end{equation}
where $Y_s$ is usually called the {\em market price of risk}.
According to Girsanov's theorem $W^\PP_t=W_t^\Q-\int_0^t Y_s ds$
is a $\PP$-Brownian motion, see~\citeN[Section
3.5]{Karshr} for details on absolutely continuous measure
transformations. Using these relations one can write the SDE for
$X$ under the physical measure $\PP$:
\begin{align}
dX_t &= (\mu(t,X_t)+\sigma(t,X_t)Y_t)dt+\sigma(t,X_t)dW^\PP_t.
\label{sdep}\end{align}

\subsection{Affine term structure models}
Affine term structure models (ATSM's) are examples of short rate
models and were introduced by \citeN{dk96}. We will give an
overview here and explicitly point out where the Feller conditions
are needed.

In an ATSM the short rate $r$ is an affine function of $X$, i.e.\
$r=\delta_0+\delta^\top X$ for some $\delta_0\in\R$,
$\delta\in\R^n$, and $X$ satisfies under $\Q$ an $n$-dimensional
affine square root SDE
\begin{equation}\label{SDE}
dX_t = (a  X_t + b)dt +\Sigma \sqrt{v(X_t)} dW^\Q_t.
\end{equation}
Here $W^\Q$ is an $n$-dimensional Brownian motion, $v(X_t)$ is a
diagonal matrix with on its diagonal the elements of the vector
\begin{equation} \mbox{diag}[v(X_t)]=\alpha+\beta X_t, \label{eq:v} \end{equation} with $\alpha
\in\RR^{n\times 1}$, $\beta\in\RR^{n\times n}$  (so
$v_{ii}(x)=\alpha_i+\beta_i x$, with $\beta_i$ the $i$-th row
vector of $\beta$). We will call these elements \emph{volatility
factors} and we write $V_t:=v(X_t)$ and $V_{i,t}:=v_i(X_t)$. For
brevity, we denote by $\sqrt{V_t}$ the matrix with on the diagonal
the square roots $\sqrt{V_{i,t}\vee 0}$, that is the square root of the maximum of $V_{i,t}$ and 0. We will also use the
notation $V_t\vee 0$ for the diagonal matrix with elements
$V_{i,t}\vee 0$. Notice that $(\sqrt{V_t})^2=V_t\vee 0$.

Since the diffusion function $x\mapsto\Sigma\sqrt{v(x)}$ is not
Lipschitz continuous for those $x$ for which $v(x)=0$, we
cannot apply standard results to assure existence and uniqueness
of a strong solution for (\ref{SDE}), unless $V_{t}>0$ almost
surely. The multivariate Feller conditions as given in
\citeN{dk96} and treated in section \ref{feller} are sufficient
for this. Thus we have encountered the first reason to impose the
Feller conditions.

In affine term structure models it is often desired that the
process $X$ also satisfies an affine square root SDE under the
physical measure $\PP$, which considerably restricts the choice
for the market price of risk $Y$. We only consider the so-called
\emph{completely} affine model, which means that we take $Y_t =
\sqrt{V_t} \lambda $ with $\lambda\in\RR^n$, and we refer to
\citeN{duf02} for other options. In this case the SDE (\ref{sdep})
takes the form
\begin{align}
dX_t &= (a  X_t + b+\Sigma(\sqrt{V_t})^2\lambda)dt+ \Sigma
\sqrt{V_t} dW^\PP_t\label{eq:sdeP}.
\end{align}
Under the condition that $V_t>0$ (elementwise on the diagonal), it
holds that $(\sqrt{V_t})^2=V_t$ and Equation~(\ref{eq:sdeP})
reduces to
\begin{align}
dX_t &= (\widehat{a}X_t+\widehat{b})dt + \Sigma\sqrt{V_t}
dW^\PP_t,\label{eq:sdehat}
\end{align}
with $\widehat{a} =a+\Sigma(\beta\odot\lambda)$, $\widehat{b}
=b+\Sigma(\alpha\odot\lambda)$.\footnote{The Hadamard-product
$\odot$ denotes entry-wise multiplication, i.e.\ $(v\odot w)_{ij}
=v_{ij} w_{ij}$ for $m\times n$-matrices $v$ and $w$. We
abbreviate $v\odot v$ to $v^{\odot 2}$. Furthermore we use the
Hadamard-product also when $v$ is an $m$-dimensional vector
(instead of an $m\times n$-matrix) and $w$ an $m\times n$-matrix
(and vice versa), so $(v\odot w)_{ij}=v_i w_{ij}$ (or $(v\odot
w)_{ij}=v_{ij} w_{i}$ in the other case).} The affine structure of
(\ref{eq:sdehat}) is thus valid under the Feller conditions, which
provides the second reason to have it imposed.

The third reason for imposing the Feller conditions is of more
practical importance, as it concerns the closed-form expression
for the bond price. Indeed, under these conditions it is possible
to solve the term structure equation (\ref{eq:pde}) by
\begin{equation}\label{eq:F}
 F(t,x) = \exp(A(T-t)+B(T-t)^\top x),
\end{equation}
for $t\in[0,T]$ and $x\in\mathcal{D}:=\{x\in\RR^n:v_i(x)>0,\forall
i\}$, where $A$ and $B$ satisfy the Riccati ordinary differential
equations (ODE's)
\begin{align}
A'&=b^\top  B + \frac{1}{2} \alpha^\top (\Sigma^\top  B)^{\odot 2}-\delta_0,\quad A(0)=0;\label{eq:A}\\
B'&= a^\top  B + \frac{1}{2} \beta^\top(\Sigma^\top  B)^{\odot
2}-\delta,\quad B(0)=0\label{eq:B},
\end{align}
see~\citeN{dk96}. However, if the bond price $D_{t,T}$ equals
$F(t,X_t)$, then it is necessary that $X_t\in\mathcal{D}$ almost
surely, since the domain of $F$ is $[0,T]\times\mathcal{D}$. In
other words, we need that $V_{i,t}>0$ almost surely, for all $i$
and $t$.

\subsection{Multivariate Feller conditions in two
dimensions}\label{feller} We will now treat the multivariate
Feller conditions for positive volatility factors, as given in
\citeN{dk96}.
\begin{prop}[\citeN{dk96}]\label{prop:feller}
Let $X$ be a solution to the affine square root SDE~(\ref{SDE}).
Then $X_t\in\mathcal{D},\forall t\geq0$ holds almost surely under
$\Q$ if the multivariate Feller conditions hold, that is for all
$i$, $j$ we have\footnote{\rm{In \citeN{dk96} it is assumed that
$c=0$, but it is not hard to see that we can take $c>0$ as well.}}

\begin{align}
\beta_i\Sigma^j & = 0 \mbox { or
$v_i=v_j+c$ for some $c\geq0$}, \label{eq:feller1} \\
\beta_i(a x+b) & >\frac{1}{2}\beta_i\Sigma\Sigma^\top\beta_i^\top
\mbox{ for all $x\in\partial\mathcal{D}^i$}. \label{eq:feller2}
\end{align}
where we write $\mathcal{D}:=\{x\in\RR^n:v_i(x)>0,i=1,\ldots,n\}$,
$\partial\mathcal{D}^i :=\{x\in\RR^n:v_i(x)=0,v_j(x)\geq0,\forall j\neq i\}$
and $\Sigma^j$ for the $j$-th column vector of $\Sigma$.
\end{prop}
In a pure latent variable model the Feller conditions can be
imposed by assuming a \emph{canonical form} for SDE (\ref{SDE}),
as shown by \citeN{ds00}. In such a canonical form the volatility
factors are equal to some of the state factors. However, we cannot
do this for macro-finance models as requiring one of the factors
to be the volatility factor is overly restrictive. Therefore we
need to extract explicit parameter restrictions from the Feller
conditions in order to impose it for a non-latent variable model.
We will do this for dimension~2.

Let $X$ be a solution to the SDE (\ref{SDE}) for $n=2$. We
distinguish three cases: {\em proportional (linear dependent)}
volatilities, {\em linearly dependent but non-proportional}
volatilities and {\em linearly independent volatilities}. The
first case is characterized by $v_2=k v_1$ for some $k \geq 0$,
the second case corresponds to $v_2=k v_1 + c$ with $k\geq 0$,
$c>0$ and in the third case one has $\det\beta\not=0$. For the
first and second case we take $k=1$, i.e.\ $k$ is absorbed in
$\Sigma$, so that we can apply the above proposition. For future
reference we also introduce $\gamma_1 =(-\beta_{12},\beta_{11})$
and $\gamma_2 = (\beta_{22},-\beta_{21})$, where the $\beta_{ij}$
are the elements of the matrix $\beta$.

\begin{description}
\item[Proportional volatilities:] Since we take $v_2=k v_1$ with
$k=1$, condition~(\ref{eq:feller1}) is automatically satisfied.
Hence, for Proposition~\ref{prop:feller} to hold, we only have to
impose~(\ref{eq:feller2}). Note that
$x\in\partial\mathcal{D}^1=\partial\mathcal{D}^2$ if and only if $
x=-\frac{\alpha_1}{|\beta_1|^2}\beta_1^\top +
y\gamma_1^\top,\mbox{ for some }y\in\RR, $ where $|\beta_1|$
denotes the Euclidean norm of the vector $\beta_1$.
Equation~(\ref{eq:feller2}) for $i=1$ becomes $
\beta_1(-\frac{\alpha_1}{|\beta_1|^2}a\beta_1^\top +
ya\gamma_1^\top+b)> \half\beta_1\Sigma\Sigma^\top \beta_1^\top
\mbox{ for all }y\in\RR. $ This reduces to the following set of
conditions
\begin{align}
-\frac{\alpha_1}{|\beta_1|^2}\beta_1a\beta_1^\top+\beta_1b &  >
\half\beta_1\Sigma\Sigma^\top \beta_1^\top\label{eq:pv1}
\\
\beta_1a\gamma_1^\top & = 0.\label{eq:pv2}
\end{align}

\item[Dependent but unproportional volatilities:]

In this case $v_2=v_1+c$, with $c>0$. Then condition
(\ref{eq:feller1}) is automatically satisfied for $i=2$, $j=1$,
but for $i=1$, $j=2$ we have to impose the extra condition
\begin{equation}
\beta_1\Sigma^2=0.\label{eq:dv}
\end{equation}
Note that $\partial\mathcal{D}^2=\emptyset$, so for
condition~(\ref{eq:feller2}) we only have to consider the case
$i=1$. The analysis is completely the same as for the case of
proportional volatilities. Hence  the conditions of
Proposition~\ref{prop:feller} are equivalent to the set of
conditions (\ref{eq:pv1}), (\ref{eq:pv2}) and (\ref{eq:dv}).

\item[Independent volatilities:]

Suppose $\det\beta\not=0$, then $\beta^{-1}$ exists. Obviously
neither $v_2=v_1+c$ nor $v_1=v_2+c$ holds true for some positive
$c$, so for condition (\ref{eq:feller1}) to hold, we need to
impose the restrictions
\begin{align}
\beta_1\Sigma^2=0, \label{eq:iv1} \\
\beta_2\Sigma^1=0. \label{eq:iv2}
\end{align}
Note that $x\in\partial\mathcal{D}^1$, respectively
$x\in\partial\mathcal{D}^2$, if and only if $\alpha+\beta
x\in\{0\}\times\RR_{\geq0}$ respectively $\alpha+\beta
x\in\RR_{\geq0}\times\{0\}$. Hence condition (\ref{eq:feller2}) is
satisfied if and only if
\begin{align}\label{cond3}
\beta_1\left(a \beta^{-1} \left(\begin{pmatrix}
    0\\
    w
  \end{pmatrix}-\alpha\right)+b\right)& > \half\beta_1\Sigma\Sigma^\top\beta_1^\top,\quad\textnormal{for
  all }w\geq 0, \\
  \beta_2\left(a \beta^{-1} \left(\begin{pmatrix}
    v\\
    0
  \end{pmatrix}-\alpha\right)+b\right)& > \half\beta_2\Sigma\Sigma^\top\beta_2^\top,\quad\textnormal{for
  all }v\geq0.\label{cond4}
\end{align}
Since
\[
\beta^{-1}=
  \frac{1}{\det\beta}\begin{pmatrix}
    \beta_{22} & -\beta_{12} \\  \\
    -\beta_{21} & \beta_{11}
  \end{pmatrix}
  =
  \frac{1}{\det\beta}\begin{pmatrix}
    \gamma_2^\top & \gamma_1^\top
  \end{pmatrix},
\]
we can reduce the restrictions (\ref{cond3}) and (\ref{cond4}) to
\begin{align*}
w\frac{\beta_1 a\gamma_1^\top}{\det\beta} +\beta_1b-\beta_1
a\beta^{-1}\alpha& > \half\beta_1\Sigma\Sigma^\top\beta_1^\top,\quad\textnormal{for all } w\geq0.\\
v\frac{\beta_2 a\gamma_2^\top}{\det\beta}  +\beta_2b-\beta_2
a\beta^{-1}\alpha& >
\half\beta_2\Sigma\Sigma^\top\beta_2^\top,\quad\textnormal{for all
} v\geq0.
\end{align*}
These hold true if and only if
\begin{align}
   \frac{\beta_1 a \gamma_1^\top}{\det\beta}&\geq 0 \label{eq:iv3} \\
      \frac{\beta_2 a \gamma_2^\top}{\det\beta}&\geq 0   \label{eq:iv4}\\
\beta_1b-\beta_1
a\beta^{-1}\alpha& > \half\beta_1\Sigma\Sigma^\top\beta_1^\top  \label{eq:iv5}\\
\beta_2b-\beta_2 a\beta^{-1}\alpha& >
\half\beta_2\Sigma\Sigma^\top\beta_2^\top.  \label{eq:iv6}
\end{align}
The first two are necessary since $v$ and $w$ can be chosen
arbitrarily large, while the latter two follow by choosing
$v=w=0$. In conclusion we can say that the requirements in
Proposition~\ref{prop:feller} are met, if the conditions
(\ref{eq:iv1}), (\ref{eq:iv2}) and (\ref{eq:iv3})-(\ref{eq:iv6})
hold.
\end{description}
It is worth noting that $X_t\in\mathcal{D},\forall t\in [0,T]$
holds almost surely under $\Q$ if and only if it holds almost
surely under $\PP$, by the equivalence of $\Q$ and $\PP$.
Furthermore, $X$ solves (\ref{SDE}) if and only if it solves
(\ref{eq:sdeP}), and under the conditions of the proposition it
also solves (\ref{eq:sdehat}). Hence one can rephrase the
conditions of the proposition by using the parameters of
(\ref{eq:sdehat}) instead of those of (\ref{SDE}), which gives the
alternative to (\ref{eq:feller2}), but {\em under}
(\ref{eq:feller1}) equivalent, condition
\begin{align}
\beta_i(\widehat{a} x+\widehat{b})
>\frac{1}{2}\beta_i\Sigma\Sigma^\top\beta_i^\top \mbox{ for all
$x\in\partial\mathcal{D}^i$}.\label{eq:Pfeller2}
\end{align}
Consequently, under $\PP$ the Feller conditions are also fulfilled under restrictions (\ref{eq:pv1}) to (\ref{eq:iv6}), with $a$ and $b$ replaced by $\widehat{a}$ and $\widehat{b}$.

\section{Affine term structure models in discrete
time}\label{sec:ctod} In this section we take Equations
 (\ref{eq:cbpR}), (\ref{SDE}), (\ref{eq:sdehat}), (\ref{eq:F}) as point of
departure and transform it into their discrete time counterparts,
using the Euler method \cite{Kloeden}. Next we investigate whether
the resulting equations are consistent with each other and in
which sense the Feller conditions are necessary in this respect.

In order not to complicate notation, we assume a discretization
factor equal to one. We write $P_{n,t}$ for the bond price at time
$t$ maturing at time $t+n$ (which corresponds to $D_{t,t+n})$ in
continuous time). Basically, all continuous time formulas are
translated to discrete-time by replacing integrals by sums and
substituting $\Delta$ for $d$. By the properties of the Brownian
motion we have $\Delta W^\Q_t=W^\Q_{t+1}-W^\Q_t\sim N(0,I)$ for
each $t$, and all these increments are mutually independent.
Therefore we write $\varepsilon^\Q_{t+1}$ instead of $dW_t^\Q$,
with $\varepsilon^\Q_{t+1}$ i.i.d.\ standard normal variables
under the risk neutral measure $\Q$. For the filtration we choose
the natural filtration $\mathcal{F}_t =
\sigma(\varepsilon^\Q_k:k=1,\ldots, t)$. We assume that $\Q$ and
the physical measure $\PP$ on $\mathcal{F}_{t}$ are related by
$d\PP = \widetilde{L}_{t} d\Q$, with $\widetilde{L}$ the
discretized exponential process
\[
 \widetilde{L}_t=\exp(\sum_{k=0}^{t-1}  \lambda^\top \sqrt{V_k} \varepsilon^\Q_{k+1}-\frac{1}{2}\sum_{k=0}^{t-1} \lambda^\top (V_k\vee 0)\lambda ).
\]
In the continuous time case, $W^\PP$ defined by $dW^\PP = dW^\Q_t
-\sqrt{V_t}\lambda dt$ is a Brownian motion under $\PP$ according to
Girsanov's theorem. Analogously, in discrete time, one can show that
the $\varepsilon^\PP_t$ defined by
$\varepsilon^\PP_{t+1}=\varepsilon^\Q_{t+1}-\sqrt{V_t}\lambda$ are
i.i.d.\ standard normal variables under $\PP$. So, we replace
$dW^\PP$ in~(\ref{eq:sdehat}) with $\varepsilon^\PP$.

Application of the above substitutions enables us to transform the
continuous time model into a discrete time model. The two
stochastic differential equations~(\ref{SDE})
and~(\ref{eq:sdehat}) under $\Q$ and $\PP$ respectively, transform
into
\begin{align}
X_{t+1}&=(I+a)X_t+b+\Sigma \sqrt{V_t}\varepsilon^\Q_{t+1}\label{Rob}\\
X_{t+1}&=(I+\widehat{a}) X_t +\widehat{b} + \Sigma
\sqrt{V_t}\varepsilon^\PP_{t+1}.\label{Prob}
\end{align}
The bond price formula (\ref{eq:cbpR})  becomes
 \begin{align}
 P_{n,t} &=\EQ(\exp(-\sum_{k=0}^{n-1} r_{t+k})|\mathcal{F}_t).\label{eq:bpR}
\end{align}
Finally, the closed form expression (\ref{eq:F}) for the bond
price in continuous time corresponds to
$\widetilde{F}(n,x)=\exp(A_n+B_n^\top x)$ in discrete time, with
$n=T-t$ and $A_n$ and $B_n$ the Euler discretizations of the
solutions of the ODE's (\ref{eq:A}) and (\ref{eq:B}). The latter
means that $A_n$ and $B_n$ satisfy the \emph{Riccati recursions}
\begin{align}
A_{n+1}&=A_n+b^\top  B_{n} + \frac{1}{2} \alpha^\top (\Sigma^\top  B_{n})^{\odot 2}-\delta_0, \qquad A_0=0;\label{eq:dA}\\
 B_{n+1}&= (I+a)^\top  B_{n} + \frac{1}{2} \beta^\top(\Sigma^\top
B_{n})^{\odot 2}-\delta, \qquad B_0=0,\label{eq:dB}
\end{align}
which are equivalent to
\begin{align}
 A_{n+1}&=A_n+(\widehat{b}-\Sigma(\alpha\odot\lambda))^\top  B_n + \frac{1}{2} \alpha^\top (\Sigma^\top
   B_n)^{\odot 2}-\delta_0,\quad &A_0&=0;\label{ricA}\\
B_{n+1}&= (I+\widehat{a}-\Sigma(\beta\odot\lambda))^\top B_n +
\frac{1}{2} \beta^\top(\Sigma^\top B_n)^{\odot 2}-\delta,\quad
&B_0&=0.\label{ricB}
\end{align}

Now that we have derived the discrete time equations, it is
important to note that it is impossible to prevent the volatility
factors $V_{t,i}$ from becoming negative, since the noise
variables are normally distributed. So in this respect it is
useless to impose the Feller conditions. Does the possibility of
negative volatility factors lead to any consistency problems for
our discrete time model? We saw that in continuous time there were
three reasons to have positive volatility factors. The first
reason concerned existence and uniqueness of a strong solution for
the SDE, which is not an issue in discrete time. The second reason
was for writing $(\sqrt{V_t})^2$ as $V_t$, which enabled us to
write (\ref{eq:sdeP}) as (\ref{eq:sdehat}), an affine square root
SDE under the physical measure. Since in discrete time $V_t$ can
always become negative, in this case it does \emph{not} hold true
that $(\sqrt{V_t})^2=V_t$, which implies that the dynamics of $X$
given by (\ref{Rob}) and (\ref{Prob}) are not consistent with each
other. Recalling the definition of $\varepsilon^\PP$, we see that
there are two possibilities to solve this problem, either we
keep~(\ref{Rob}) and replace~(\ref{Prob}) by
\begin{align}
X_{t+1}&=(I+a) X_t +b + \Sigma (V_t\vee0)\lambda + \Sigma
\sqrt{V_t}\varepsilon^\PP_{t+1},\label{eq:newProb}
\end{align}
or we keep~(\ref{Prob}) and replace~(\ref{Rob}) by
\begin{align}
X_{t+1}&=(I+\widehat{a}) X_t +\widehat{b} - \Sigma (V_t\vee0)\lambda
+ \Sigma \sqrt{V_t}\varepsilon^\Q_{t+1}.\label{eq:newRob}
\end{align}
We opt for the latter, because an attractive expression under the
physical measure is preferable in view of estimation of the
parameters.

\begin{figure}
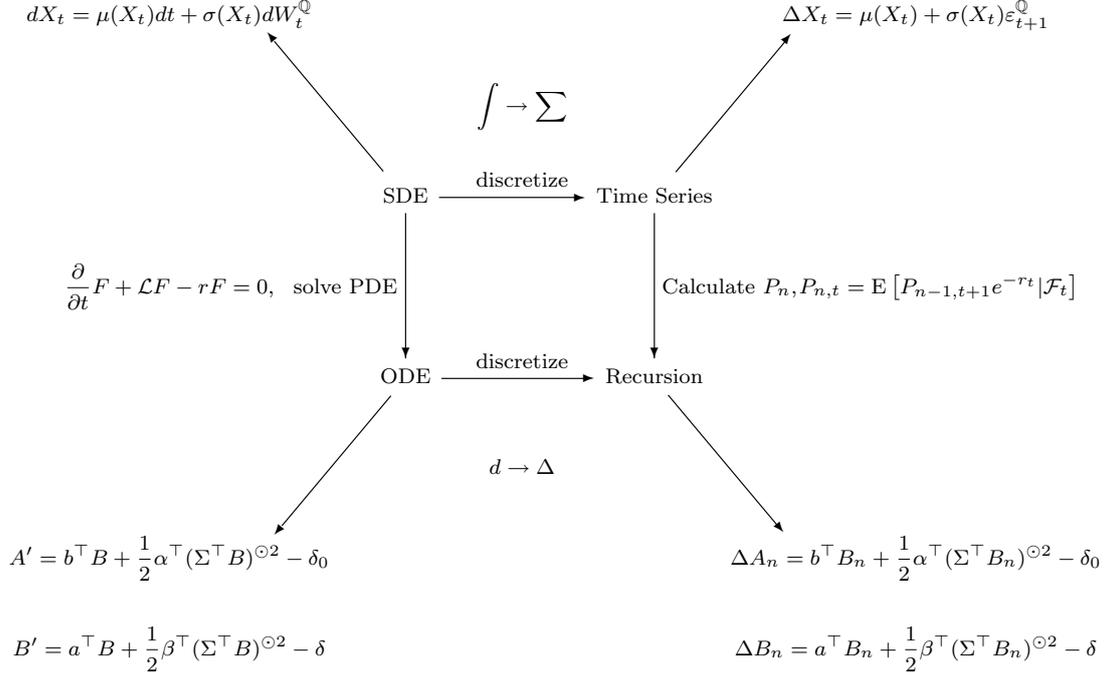

 {\scriptsize  \begin{diagram}[fixed,balance,height=1.2cm,width=1cm]
dX_t=\mu(X_t)dt+\sigma(X_t)dW^\Q_t &       &                               &                              &                                   &     &\Delta X_{t}=\mu(X_t)+\sigma(X_t)\varepsilon^\Q_{t+1} \\
                                &\luTo  &                               & \int\to\sum            &                                   &\ruTo&\\
                                &       &\textnormal{SDE}               &\rTo^{\textnormal{discretize}} &\textnormal{Time Series}          &\\
 \frac{\partial}{\partial t}F+\mathcal{L}F-rF=0,                               &       &\dTo^{\textnormal{solve PDE}}  &                               &\dTo_{\textnormal{Calculate $P_n$,}}&&\qquad P_{n,t}=\E\left[P_{n-1,t+1}e^{-r_t}|\mathcal{F}_t\right] \\
                                &       &\textnormal{ODE}               &\rTo^{\textnormal{discretize}} &\textnormal{Recursion}   &\\
                                &\ldTo  &                               &  d\to\Delta                     &                                   &\rdTo  &\\
A'=b^\top  B + \frac{1}{2} \alpha^\top (\Sigma^\top  B)^{\odot 2}-\delta_0  &&&&&&
\Delta A_{n}=b^\top  B_{n} + \frac{1}{2} \alpha^\top (\Sigma^\top  B_{n})^{\odot 2}-\delta_0\\
B'= a^\top  B + \frac{1}{2} \beta^\top(\Sigma^\top  B)^{\odot
2}-\delta     &&&&&& \Delta B_{n}= a^\top  B_{n} + \frac{1}{2}
\beta^\top(\Sigma^\top  B_{n})^{\odot 2}-\delta
\end{diagram}}
\caption{Commutative diagram for calculating the bond-price in discrete time} \label{comdiag}
\end{figure}

The third reason why we needed positive volatility factors in
continuous time was to solve the term structure equation
(\ref{eq:pde}) in order to obtain a closed form expression $F$ for
the bond price. There is no discrete time analogue of a term
structure equation. However, using induction and the properties of
a log-normal variable, we can \emph{algebraically} derive a closed
form expression for the bond price in discrete time, and, just as
in continuous time, we need positiveness of the volatility factors
for this, see Proposition~\ref{prop:dtaffine} below. Remarkable is
that this leads to the same expression as $\widetilde{F}$, the
discretization of the closed form expression $F$ in continuous
time. This is summarized in Figure~\ref{comdiag} in a {\em
commutative diagram}.

\begin{prop}\label{prop:dtaffine}
Let $X$ satisfy (\ref{Rob}). Then for $P_{n,t}$ given by
(\ref{eq:bpR}) it holds that
\begin{equation}\label{eq:pnt}
P_{n,t}\geq \widetilde{F}(n,t)=\exp(A_n+B_n^\top X_t),
\end{equation}
with equality iff $V_t\geq0$ almost surely for all $t$. The
scalars $A_n$ and vectors $B_n$ satisfy the Riccati recursions
(\ref{eq:dA}) and (\ref{eq:dB}).
\end{prop}
\begin{proof}
We give a proof by induction. For $n=0$ it holds that
$P_{n,t}=P_{0,t}=1$, so the statement holds true with $A_0=B_0=0$.
Now suppose $P_{n-1,t}\geq\exp(A_{n-1}+B_{n-1}^\top X_t)$ for all
$t$ and for a certain $n\in\mathbb{N}$. We write
\[
P_{n,t}=\E_\Q\left[\exp\left(\left.-\sum_{k=0}^{n-1}
r_{t+k}\right)\right|\mathcal{F}_t\right]
=\E_\Q\left[P_{n-1,t+1}e^{-r_t}|\mathcal{F}_t\right],
\]
and use the induction hypothesis to get
\begin{align*}
P_{n,t}&\geq\E_\Q[\exp(A_{n-1}+B_{n-1}^\top X_{t+1}-\delta_0
-\delta^\top
X_t)|\mathcal{F}_t]\\
&=\E_\Q[\exp(A_{n-1}+B_{n-1}^\top ((I+a)
X_t+b+\Sigma\sqrt{V_t}\varepsilon^\Q_{t+1})-\delta_0-\delta^\top X_t)|\mathcal{F}_t]\\
&=\exp(A_{n-1}+B_{n-1}^\top b-\delta_0+((I+a)^\top
B_{n-1}-\delta)^\top X_t)
\E_\Q[\exp(B_{n-1}^\top\Sigma\sqrt{V_t}\varepsilon^\Q_{t+1})|\mathcal{F}_t]\\
&\geq\exp(A_{n-1}+B_{n-1}^\top b-\delta_0+((I+a)^\top
B_{n-1}-\delta)^\top
X_t+\frac{1}{2}B_{n-1}^\top\Sigma V_t\Sigma^\top B_{n-1})\\
&=\exp\bigg(\underbrace{A_{n-1}+B_{n-1}^\top b-\delta_0+\frac{1}{2}\alpha^\top (\Sigma^\top  B_{n-1})^{\odot 2}}_{\displaystyle A_n}\\
&\qquad\qquad\qquad\qquad\qquad\qquad+\underbrace{((I+a)^\top
B_{n-1}-\delta+\frac{1}{2}\beta^\top(\Sigma^\top B_{n-1})^{\odot
2}}_{\displaystyle B_n})^\top X_t\bigg),
\end{align*}
where we have used that (with $\ell=\Sigma^\top  B_{n-1}$)
\begin{equation}\label{eq:ineq}
\E_\Q[\exp(\ell^\top
\sqrt{V_t}\varepsilon^\Q_{t+1})|\mathcal{F}_t]=\exp(\half
\ell^\top (V_t\vee 0) \ell)\geq\exp(\half \ell^\top V_t \ell)
\end{equation}
and
\begin{align*}
\ell^\top V_t \ell &= \ell^\top(\alpha\odot \ell) +\ell^\top(\beta X_t\odot \ell)=\alpha^\top (\ell\odot \ell)+(\beta X_t)^\top
(\ell\odot \ell)= \alpha^\top \ell^{\odot2}+X_t^\top\beta^\top \ell^{\odot2}\\&= \alpha^\top \ell^{\odot2}+(\beta^\top
\ell^{\odot2})^\top X_t.
\end{align*}
If $V_t\geq0$ then the inequality in (\ref{eq:ineq}) is an
equality. This proves the assertion.
\end{proof}

Of course, the probability that $V_t$ gets negative is positive
for all $t$, so $P_{n,t}$ is \emph{not} equal to
$\widetilde{F}(n,X_t)$. Note also that the above proposition is
inapplicable for our model, as $X$ solves (\ref{eq:newRob})
instead of (\ref{Rob}). However, from a heuristic point of view,
if $V_t$ gets negative with very small probability,
(\ref{eq:newRob}) and (\ref{Rob}) are ``almost'' equivalent and
the inequality in (\ref{eq:ineq}) is ``almost'' an equality. This
suggests that $\widetilde{F}$ might be a good approximation for
$P$.

A priori though, it is not clear whether a model with parameters
that satisfy the Feller conditions yields a better approximation
than a model without parameter restrictions, as we do not know how
the Feller conditions affect the probability for negative $V_t$ in
discrete time. In fact, in the introduction we already saw an
example in discrete time in which the relative frequency
of negative $V_t$ was rather small, without having the Feller
conditions imposed.

In the remaining sections we implement and estimate the discrete
time model for dimension 2 using real data and we investigate how
well in this case $\widetilde{F}$ approximates $P$. This is done by
comparing $\widetilde{F}$ with Monte Carlo computations for the bond
price $P$, based on (\ref{eq:bpR}) and (\ref{eq:newRob}).

\section{Implementation and estimation of the discrete time ATSM} \label{sec:model}

In this section, we investigate a two-factor model with the {\em
ex-ante real short term rate} and {\em expected inflation} as state
variables, the nominal short rate  being the sum of these two
factors. The interaction between interest rates and inflation is
important, for instance for pension funds, as most of them have the
intention to give indexation, whereas index linked bonds are hardly
available. For the Netherlands, another important motivation for
modeling this link is that supervision is geared toward nominal
guarantees.

\subsection{Specification of the model}

Let  $X_{1,t}$ denote the ex-ante real short term rate at time $t$
and $X_{2,t}$ the expected inflation. We use our dynamic model for
quarterly data, so time is measured in quarters. Consequently, $r_t$
as used in the pricing formulas for bonds, is given in ordinary
fractions per time unit, in our case per quarter. For numerical and
readability reasons however, we want to express $X$ in percentages
per year. Therefore, we have $r_t=(X_{1,t} + X_{2,t})/400$.

With respect to inflation, we are primarily interested in the
ex-ante expectation and not so much in past realizations. Let
$\pi_{t+1}$ denote the inflation rate from $t$ to $t+1$ (also in
percentages per year), and $X_{2,t}$ its ex-ante expectation at
date $t$. The observed processes are the short nominal rate $r_t$
and $\pi_t$. Since the inflation rate exhibits a seasonal pattern,
we also include a seasonal contribution $S_t$ in the model.

Apart from the dynamics of the state process $X$ as given in
Equation~(\ref{Prob}), our model is described by the following
equations, that relate the state variables to the observations.
\begin{subequations}\label{eq:sdynamics}
    \begin{align}
      r_t &= (X_{1,t} + X_{2,t})/400 \label{eq:r}\\
        \pi_{t+1} &= X_{2,t}  + S_{t+1} + \omega_\pi \sqrt{V_{2,t-1}}\xi_{\pi,t+1} \label{eq:inflation} \\
        S_{t+1}   &= - S_t - S_{t-1} - S_{t-2} + \omega_s \sqrt{V_{2,t-1}}\xi_{s,t+1}, \label{eq:s}
    \end{align}
\end{subequations}
where  $\xi_{\pi,t}$ and $\xi_{s,t}$ are standard normally
distributed error terms, that are independent under the physical
measure of $\varepsilon_t^\PP$ (which is the error term in the
equation for $X_t$).

The data we are using for estimation are the observed longer
maturity yields (denoted $r_{n,t}$, measured in fractions per
quarter). These are modeled by the exponential affine expression
for the bondprice plus a measurement error, which is assumed to be
independently identically distributed among maturities:
\begin{equation}\label{eq:rn}
r_{n,t} :=  -\left(A_n+B_n X_t\right)/n + \left(\nu_0+\nu_1
\sqrt{V_{1,t-1}}+\nu_2 \sqrt{V_{2,t-1}}\right) \xi_{n,t},
\end{equation}
under the restrictions $\nu_i \geq 0$, and where $\xi_{n,t}\sim N(0,I)$.

Having fully specified  the model, we turn to the estimation
procedure. A complicating matter is that both factors are not
observed. Therefore, the extended Kalman filter \cite{har89} is
used to estimate the models.\footnote{The extension is due to the
variance equation, that includes state variables. Consequently,
the true variance process is not known exactly, but has to be
estimated as well. The resulting inconsistency does not seem to be
very important, though in short samples the mean reversion
parameters are often biased upwards, see \citeN{lun97},
\citeN{ds99}, \citeN{dej00}, \citeN{bol01}, \citeN{cs03},
\citeN{ds04}, and \citeN{der06}.} In principle, all parameters can
be estimated simultaneously. In practice however, a one-step
procedure tends to lead to unrealistic expected inflation
predictions as the best fit for the bond prices are not
necessarily achieved for the most realistic expected inflation
estimates. As an appropriate modeling of the time series dynamics
of interest rates and inflation is considered more important than
the lowest measurement error for bond prices, we prefer a two-step
procedure. In the first step, the parameters for system
(\ref{eq:sdynamics}) are estimated, combined with the
dynamics~(\ref{Prob}) for $X_t$  and its volatility~(\ref{eq:v}).
In the second step, the system is augmented by the equations for
the long-term yields (\ref{eq:rn}), and we estimate $\lambda$ and
$\nu$, using the Riccati recursions (\ref{ricA}) and (\ref{ricB}),
conditional on the first-step parameters.

\subsection{Estimation results} \label{sec:estimate}

The models are estimated with quarterly German data over the
sample period 1959 to 2007. In order to estimate the dynamics
between interest rates and inflation correctly, a long sample
period is preferred. On the other hand, prices of zero coupon
bonds are only available for a relatively short sample period,
especially for longer maturities. Therefore, an unbalanced panel
was used, with the short rate and inflation data starting in the
last quarter of 1959, 1, 2, 4, 7 and 10-year rates starting the
third quarter of 1972, the 15 year rate from 1986:II, and the
30-year rate from 1996:I on.

\begin{table}[h]
\caption{Estimation results without the Feller conditions imposed}
\label{tab:parmU} \vspace*{1mm}
\centering{\small \begin{tabular}{cccc}\hline\\
& \mbox{Proportional volatilities}& \mbox{Dependent volatilities} & \mbox{Independent volatilities} \\\hline\\
$\begin{array}{c}\left(-\widehat{a}^{-1}\widehat{b}\right)^\top
\end{array}$ & $\left[\begin{array}{cc}
2.36&3.05\\(2.9)&(4.5) \end{array}\right]$ &
$\left[\begin{array}{cc} 2.39&3.03\\(3.0)&(6.3) \end{array}\right]$ & $\left[\begin{array}{cc} 2.33&3.12\\(2.0)&(5.2) \end{array}\right]$\\\\
$I+\widehat{a}$&$\left[\begin{array}{cc}0.926&0.087\\(26.5)&(1.7)\\-0.002&0.938\\(0.1)&(24.0)\end{array}\right]$
&$\left[\begin{array}{cc}0.948&0.111\\(26.3)&(2.8)\\-0.026&0.950\\(0.9)&(25.5)\end{array}\right]$
&$\left[\begin{array}{cc} 0.946 & 0.102\\(24.4)&(2.2)\\ -0.006 & 0.940\\(0.2)&(24.3) \end{array}\right]$ \\\\
$\begin{array}{c}\alpha^\top \end{array}$ &
$\left[\begin{array}{cc} -0.377 &
-0.377\\(2.3)&(2.3)\end{array}\right]$ & $\left[\begin{array}{cc}
-0.373 & -0.165\\(12.2)&(1.4)\end{array}\right]$ &
$\left[\begin{array}{cc} -0.377 & -0.081\\(3.2)&(1.5)\end{array}\right]$\\\\
$\begin{array}{c}\beta \end{array}$ & $\left[\begin{array}{cc}
0.105 & 0.230
\\(2.3)&(2.9)\\ 0.105 & 0.230\\(2.3)&(2.9)
\end{array}\right]$ & $\left[\begin{array}{cc} 0.108 &
0.194\\(4.7)&(10.4)\\ 0.108 & 0.194\\(4.7)&(10.4)
\end{array}\right]$ & $\left[\begin{array}{cc} 0.117 &
0.193\\(2.9)&(3.3)\\ 0.015 &
0.091\\(1.4)&(2.6)\end{array}\right]$\\\\
$\Sigma$ & $\left[\begin{array}{cc} 1 & 0\\(-)&(-)\\ -0.257 &
0.639\\(1.5)&(5.3) \end{array}\right]$ &
$\left[\begin{array}{cc} 1 & -0.260\\(-)&(1.7)\\
0.052 & 0.547\\(0.5)&(4.2) \end{array}\right]$
 & $\left[\begin{array}{cc} 1 & -0.526\\(-)&(3.0)\\ 0.041 & 1\\(0.3)&(-) \end{array}\right]$\\\\
$\begin{array}{c}\lambda^\top \end{array}$ &
$\left[\begin{array}{cc} 0.105 & -0.129\\(0.7)&(1.3)
\end{array}\right]$ & $\left[\begin{array}{cc} 0.108 & -0.136 \\(0.6)&(1.4)
\end{array}\right]$ & $\left[\begin{array}{cc} 0.0524 & -0.209 \\(0.3)&(1.1)
\end{array}\right]$\\\\\hline
\end{tabular}

\vspace*{3mm} Estimation sample 1959:IV - 2007:II

Absolute two-step consistent t-values in parenthesis.}
\end{table}

Table~\ref{tab:parmU} shows the estimation results for the models
without the Feller conditions imposed. As only the conditional
covariance matrix of the noise terms, which is given by $\Sigma
(V_t\vee 0) \Sigma^\top$, is identifiable, we fix $\Sigma_{11}=1$ in
all models, we choose $\beta_1=\beta_2$ in the proportional and
dependent models, we take $\Sigma_{12}=0$ in the proportional model,
and $\Sigma_{22}=1$ in the independent model.

The mean real short rate is about 2.4\% per year, whereas the mean
inflation rate is just over 3\%, the values in the top row of
Table~\ref{tab:parmU}. With respect to the interaction between the
short real rate and expected inflation, the lagged response
($\widehat{a}$) is in accordance with economic theory. Higher
expected inflation leads to higher real rates, whereas higher real
rates depress future inflation. The latter effect is far from
significant though. With respect to volatility ($\alpha$ and
$\beta$), both higher real rates and higher expected inflation lead
to significantly higher variances.

\begin{table}[h] \caption{Estimation results with Feller conditions imposed} \label{tab:parmR} \vspace*{1mm}
\centering{\small \begin{tabular}{cccc}\hline\\
& \mbox{Proportional volatilities}& \mbox{Dependent volatilities} & \mbox{Independent volatilities} \\\hline\\
$\begin{array}{c}\left(-\widehat{a}^{-1}\widehat{b}\right)^\top
\end{array}$ & $\left[\begin{array}{cc}
2.34&3.04\\(2.4)&(4.3) \end{array}\right]$ &
$\left[\begin{array}{cc} 2.36&3.04\\(1.9)&(3.3) \end{array}\right]$ & $\left[\begin{array}{cc} 2.91&2.83\\(2.4)&(3.8) \end{array}\right]$\\\\
$I+\widehat{a}$&$\left[\begin{array}{cc}0.924&0.083\\(25.0)&(1.5)\\0.016&0.925\\(0.9)&(25.0)\end{array}\right]$
&$\left[\begin{array}{cc}0.933&0.061\\(22.2)&(1.2)\\0.021&0.936\\(1.1)&(28.7)\end{array}\right]$
&$\left[\begin{array}{cc} 0.974 & -0.009\\(52.4)&(0.4)\\ 0 & 0.958\\(-)&(47.9) \end{array}\right]$ \\\\
$\begin{array}{c}\alpha^\top \end{array}$ &
$\left[\begin{array}{cc} -0.412 &
-0.412\\(2.3)&(2.3)\end{array}\right]$ & $\left[\begin{array}{cc}
-0.098 & -0.062\\(5.7)&(1.4)\end{array}\right]$ &
$\left[\begin{array}{cc} 0.020 & -0.108\\(0.7)&(4.5)\end{array}\right]$\\\\
$\begin{array}{c}\beta \end{array}$ & $\left[\begin{array}{cc}
0.108 & 0.252\\(2.2)&(2.9)\\ 0.108 & 0.252\\(2.2)&(2.9)
\end{array}\right]$ & $\left[\begin{array}{cc} 0.028 &
0.049\\(1.5)&(1.6)\\0.028 & 0.049\\(1.5)&(1.6)
\end{array}\right]$ & $\left[\begin{array}{cc} 0.071 &
-0.044\\(2.6)&(1.2)\\ 0 & 0.100\\(-)&(5.7)\end{array}\right]$\\\\
$\Sigma$ & $\left[\begin{array}{cc}  1 & 0\\(-)&(-)\\ -0.292 &
0.640\\(1.8)&(5.6) \end{array}\right]$ &
$\left[\begin{array}{cc} 1 & -1.620\\(-)&(3.1)\\
1.116 & 0.915\\(1.9)&(2.0) \end{array}\right]$
 & $\left[\begin{array}{cc} 1 & 0.615\\(-)&(3.4)\\ 0 & 1\\(-)&(-) \end{array}\right]$\\\\
$\begin{array}{c}\lambda^\top \end{array}$ &
$\left[\begin{array}{cc} 0.0050 & -0.124 \\(0.0)&(1.1)
\end{array}\right]$ & $\left[\begin{array}{cc} -0.153 &
0.667 \\(1.3)&(1.6)
\end{array}\right]$ & $\left[\begin{array}{cc} -0.397 & -0.125 \\(1.2)&(0.1)
\end{array}\right]$\\\\\hline
\end{tabular}

\vspace*{3mm} Estimation sample 1959:IV - 2007:II

Absolute two-step consistent t-values in parenthesis.}
\end{table}

Table~\ref{tab:parmR} shows the results for the models that are
restricted to fulfill the Feller conditions. In the independent
volatility model, initially obtained estimates for $\Sigma_{21},
\beta_{21}$ and $\widehat{a}_{21}$ were practically zero. Therefore,
a zero value was subsequently imposed to increase accuracy of the
other parameters. In this model, higher inflation now has a negative
(though not significant) impact ($\hat{a}_{12}$) on future short
term interest rates, which is contrary to economic theory, whereas
in the previous case when the Feller conditions were not imposed, we
found for this coefficient a positive value. In the other models,
the impact of inflation on lagged real rates ($\widehat{a}_{21}$) is
now positive, which is also in contrast with economic theory.

\section{Monte Carlo results}\label{sec:mc}

\begin{figure}[b]
\begin{center}
\caption{Mean and 99\% confidence interval of the difference between
simulated and analytical yields if starting state variables are in
equilibrium}\label{fig:error_eq}
\includegraphics[width=\textwidth,angle=0]{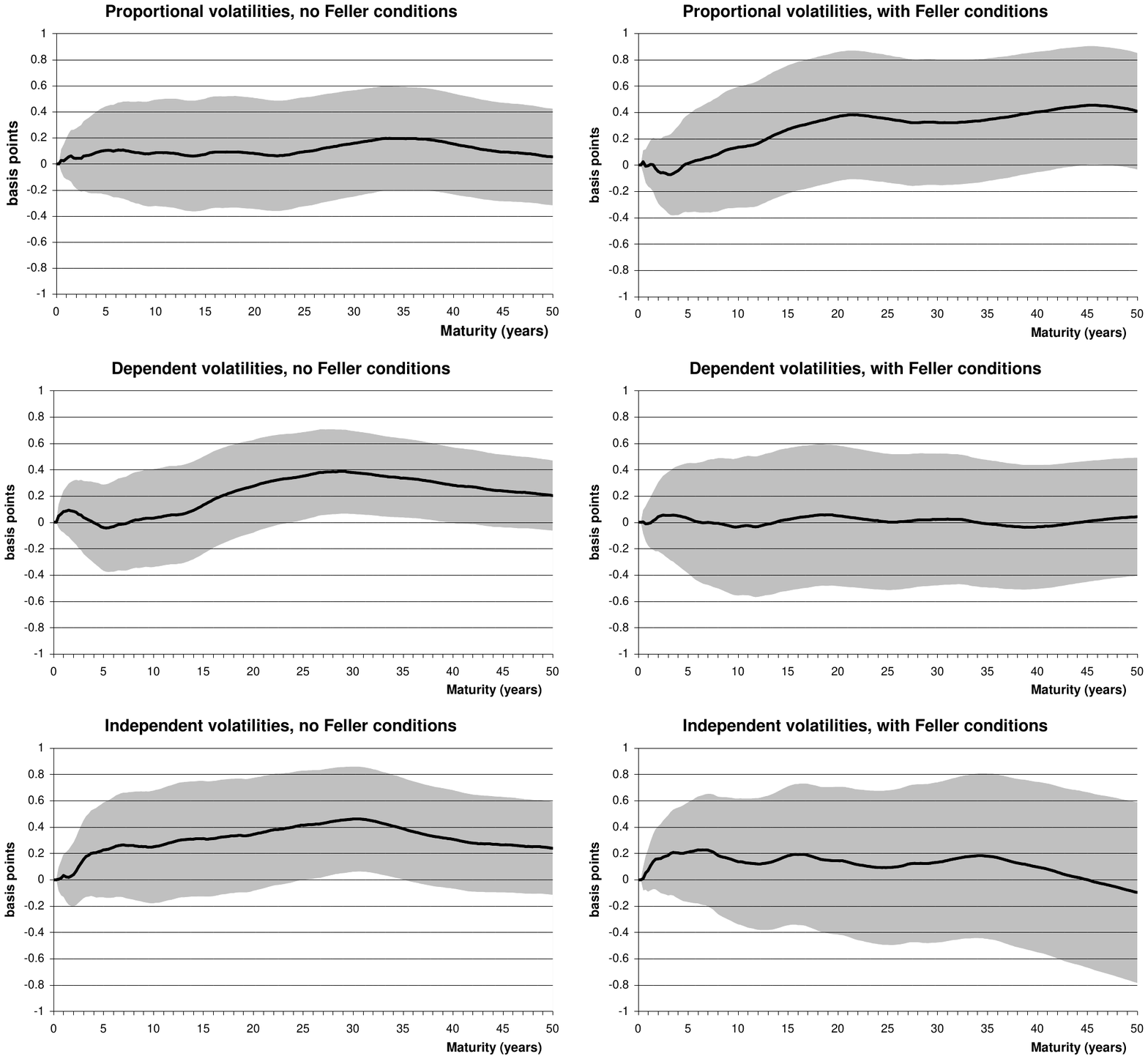}
\end{center}
\vspace*{-8.2cm}
\end{figure}

Figure \ref{fig:error_eq} shows the approximation errors made by
the analytical expressions, in terms of yields, for each of the
six cases as presented in Tables~\ref{tab:parmU}
and~\ref{tab:parmR}. The Monte Carlo simulations are based one
million sample paths (containing 200 quarters) for the state
variables. The yields are computed assuming the initial state
variables are at their equilibrium values, which were a real short
rate of about 2.4\% and expected inflation of just over 3\%.

Ignoring the fact that volatilities are cut-off at zero does not
seem to be important. Indeed, the 99\% confidence band for the
maximum approximation error in terms of yields stays within plus
and minus one basis point (0.01\%) for all models. It does not
make much difference whether the Feller conditions are imposed
(second column) or not (first column). Zero is almost always
included in the confidence band, except for some maturities for
the dependent and independent volatility models without Feller
conditions imposed. For the proportional volatility model, there
is never a problem, whether the Feller conditions are imposed or
not.

\begin{figure}[b]
\begin{center}
\caption{Mean and 99\% confidence interval of the difference
between simulated and analytical yields. Dependent volatility
model without Feller conditions, starting from a volatility of
zero.}\label{fig:error_low} \vspace*{-1.2cm}
\hspace*{-1cm}\includegraphics[width=10cm,angle=270]{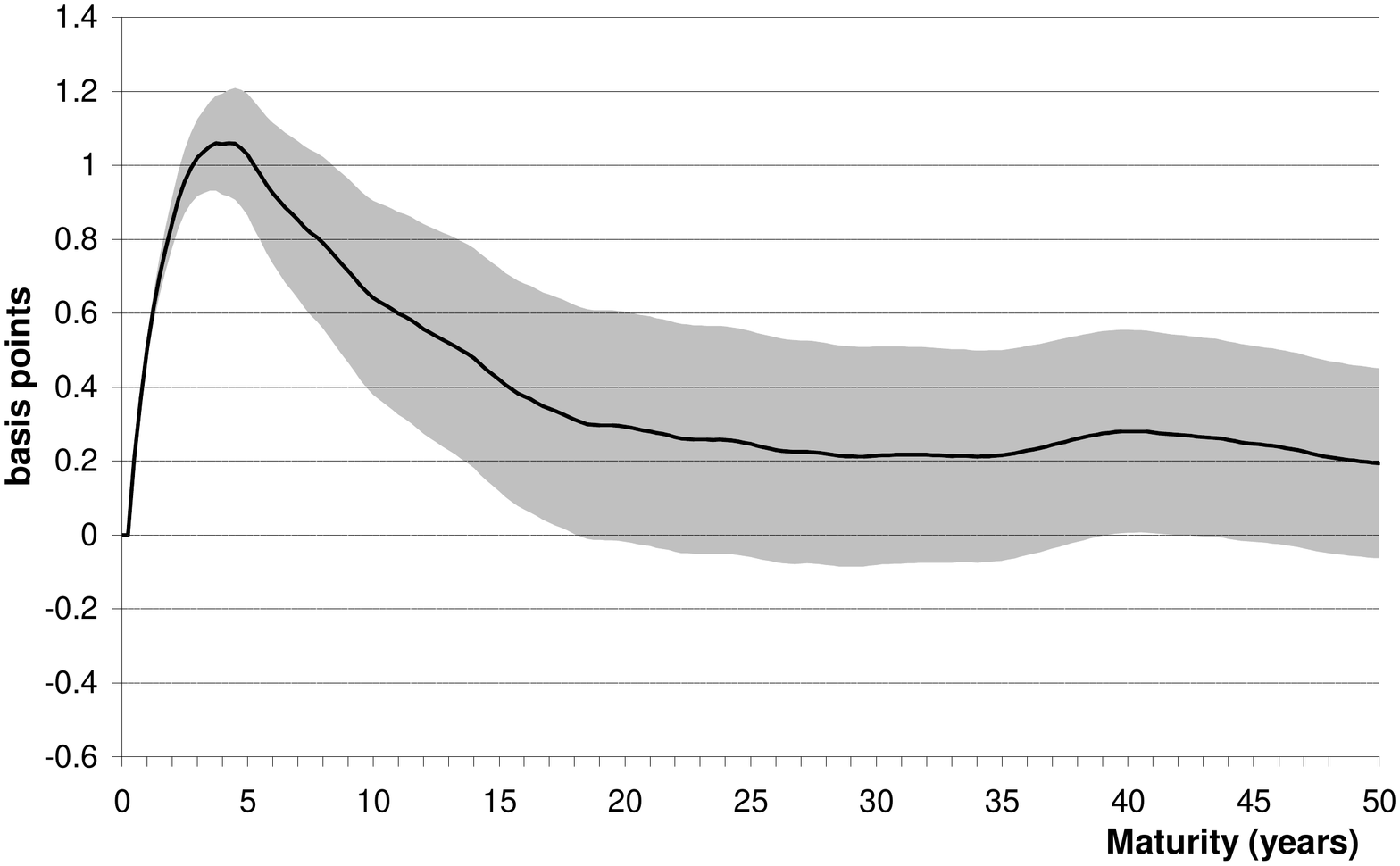}
\end{center}
\vspace*{-1.2cm}
\end{figure}

It might be the case that these good approximations are due to the
fact the approximation errors are calculated for the equilibrium
yield curve. If the initial state variables imply a volatility
closer to zero, ignoring the cut-off at $v=0$ might be more serious.
Therefore, we also calculated the approximation errors for those
state variables for which volatility was the lowest in the past.
Figure \ref{fig:error_low} shows the worst result we found. Indeed,
for maturities up to 15 years, the simulated yields are
significantly higher than the analytical ones. The reason is that
for the initial state variables, the volatility is cut-off at zero.
As the state variables evolve according to (\ref{eq:newRob}),
whereas the formulas underlying the formulas assume (\ref{Rob}),
systematic differences arise. Moreover, as the simulated yields are
almost deterministic for short maturities, the confidence band is
extremely small. In economic terms, the approximation error is still
negligible though (at most one basis point).

Finally, as the Feller conditions do not guarantee positive
volatilities in the discrete time model, imposing them does not
preclude statistically significant approximation errors from
arising either. Indeed for both the dependent and independent
volatility models with Feller conditions we found starting
conditions for which significant negative approximation errors for
maturities up to 17 years occur. However, as the maximum magnitude
of these errors is at most 0.5 basis point, the economic relevance
is again negligible.

\section{Conclusions}\label{sec:conclusions}
The Feller conditions are imposed on a continuous time multivariate
square root process in order to have well defined strong solutions
to the stochastic differential equations and to ensure that the
roots have nonnegative arguments. For a discrete time approximate
model, the Feller conditions loose part of their relevance.
Existence of strong solutions is not an issue anymore and since the
noise involves standard normal errors, there is always a positive
probability that arguments of square roots become negative.
Nevertheless, keeping in mind the idea that a discrete time model is
an approximation of a continuous time model, it is natural to still
impose the Feller conditions. On the other hand, it has also been
observed that even without the Feller conditions imposed, for a
practically relevant model, negative arguments rarely occur.

We have investigated the relevance of imposing the Feller
conditions for a two-factor affine term structure model, where the
factors (ex-ante real short term interest rate and expected
inflation) are modelled as a square root process. As we want to
allow volatilities to depend on both state variables, the models
are not estimated in canonical form. Therefore we also explicitly
presented the Feller conditions for square root models not in
canonical form.

Three different models have been used, that have been referred to
as models with proportional, dependent and independent
volatilities, either with or without the Feller conditions on the
parameters. The parameters of each of the underlying models have
been estimated using quarterly German data. The restrictions
involved in imposing the Feller conditions resulted in unappealing
economic results. In the proportional and dependent volatility
models, the restrictions imply a positive impact of interest rates
on inflation, whereas in the independent volatility model,
inflation now leads to lower interest rates. Both elements are
contrary to the accepted economic theory.

For these six cases we have compared the resulting yields, that
are either obtained by (approximate) analytic exponentially affine
expressions or those obtained through Monte Carlo simulations of
very high numbers of sample paths. It turned out that the
approximation errors in analytical yields were rarely
statistically significant, and never economically relevant, as
they were always below one basis point. In particular a
proportional volatility model without the Feller conditions
imposed already gave very good results.

\bibliographystyle{chicagocf}
\bibliography{termstruc}

\begin{thebibliography}{}

\bibitem[\protect\citeauthoryear{Ang and Bekaert}{Ang and Bekaert}{2004}]{ab04}
Ang, A. and K.G.~Bekaert (2004),
\newblock `The term structure of real rates and expected inflation',
\newblock CEPR Discussion Paper 4518.

\bibitem[\protect\citeauthoryear{Ang and Piazzesi}{Ang and
  Piazzesi}{2003}]{ap03}
Ang, A. and M.~Piazzesi (2003),
\newblock `A no-arbitrage vector autoregression of term structure dynamics with
  macroeconomic and latent variables',
\newblock {\em Journal of Monetary Economics\/}~{\bf 50\/}(4), 745--787.

\bibitem[\protect\citeauthoryear{Backus, Foresi, and
  Telmer}{Backus~\emph{et~al.}}{2001}]{bft01}
Backus, D., S.~Foresi, and C.~Telmer (2001, 02),
\newblock `Affine term structure models and the forward premium anomaly',
\newblock {\em Journal of Finance\/}~{\bf 56\/}(1), 279--304.

\bibitem[\protect\citeauthoryear{Bernanke, Reinhart, and
  Sack}{Bernanke~\emph{et~al.}}{2005}]{brs05}
Bernanke, B., V.~Reinhart, and B.~Sack (2005),
\newblock `Monetary policy alternatives at the zero bound: {A}n empirical
  assessment',
\newblock {\em Brookings Papers on Economic Activity\/}, {\bf 2}, 1--78.

\bibitem[\protect\citeauthoryear{Bolder}{Bolder}{2001}]{bol01}
Bolder, D.J. (2001),
\newblock `Affine term-structure models: {T}heory and implementation',
\newblock Working Paper 2001-15, Bank of Canada.

\bibitem[\protect\citeauthoryear{Brigo and Mercurio}{Brigo and
  Mercurio}{2006}]{brimer}
Brigo, D. and F.~Mercurio (2006),
\newblock {\em Interest rate models---theory and practice\/} (Second ed.),
\newblock Springer Finance, Berlin: Springer-Verlag,
\newblock With smile, inflation and credit.

\bibitem[\protect\citeauthoryear{Campbell and Viceira}{Campbell and
  Viceira}{2002}]{cv02}
Campbell, J.Y. and L.M.~Viceira (2002),
\newblock {\em Strategic Asset Allocation},
\newblock Clarendon Lectures in Economics: Oxford University Press.

\bibitem[\protect\citeauthoryear{Chen and Scott}{Chen and Scott}{2003}]{cs03}
Chen, R.R. and L.~Scott (2003),
\newblock `Multi-factor {C}ox-{I}ngersoll-{R}oss models of the term structure:
  {E}stimates and tests from a {K}alman filter model',
\newblock {\em Journal of Real Estate Finance and Economics\/}~{\bf 27\/}(2),
  143--172.

\bibitem[\protect\citeauthoryear{Dai, Le, and
  Singleton}{Dai~\emph{et~al.}}{2005}]{dls05}
Dai, Q., A.~Le, and K.J.~Singleton (2005),
\newblock `Discrete-time dynamic term structure models with generalized market
  prices of risk',
\newblock Manuscript, Stanford University.

\bibitem[\protect\citeauthoryear{Dai and Singleton}{Dai and
  Singleton}{2000}]{ds00}
Dai, Q. and K.J.~Singleton (2000),
\newblock `Specification analysis of affine term structure models',
\newblock {\em Journal of Finance\/}~{\bf 55\/}(5), 1943--1978.

\bibitem[\protect\citeauthoryear{{De Jong}}{{De Jong}}{2000}]{dej00}
{De Jong}, F. (2000),
\newblock `Time-series and cross-section information in affine term structure
  models',
\newblock {\em Journal of Business and Economic Statistics\/}, {\bf 18},
  300--314.

\bibitem[\protect\citeauthoryear{{De Rossi}}{{De Rossi}}{2006}]{der06}
{De Rossi}, G. (2006),
\newblock `Unit roots and the estimation of multifactor
  {C}ox-{I}ngersoll-{R}oss models',
\newblock Manuscript, Cambridge University.

\bibitem[\protect\citeauthoryear{Dewachter and Lyrio}{Dewachter and
  Lyrio}{2006}]{dl06}
Dewachter, H. and M.~Lyrio (2006),
\newblock `Macro factors and the term structure of interest rates',
\newblock {\em Journal of Money, Credit, and Banking\/}~{\bf 38\/}(1),
  119--140.

\bibitem[\protect\citeauthoryear{Dewachter, Lyrio, and
  Maes}{Dewachter~\emph{et~al.}}{2004}]{dlm04}
Dewachter, H., M.~Lyrio, and K.~Maes (2004),
\newblock `The effect of monetary unification on {G}erman bond markets',
\newblock {\em European Financial Management\/}~{\bf 10\/}(3), 487--509.

\bibitem[\protect\citeauthoryear{Dewachter, Lyrio, and
  Maes}{Dewachter~\emph{et~al.}}{2006}]{dlm06}
Dewachter, H., M.~Lyrio, and K.~Maes (2006),
\newblock `A joint model for the term structure of interest rates and the
  macroeconomy',
\newblock {\em Journal of Applied Econometrics\/}, {\bf 21}, 439--462.

\bibitem[\protect\citeauthoryear{Duan and Simonato}{Duan and
  Simonato}{1999}]{ds99}
Duan, J.C. and J.G.~Simonato (1999),
\newblock `Evaluating an alternative risk preference in affine term structure
  models',
\newblock {\em Review of Quantitative Finance and Accounting\/}, {\bf 13},
  111--135.

\bibitem[\protect\citeauthoryear{Duffee}{Duffee}{2002}]{duf02}
Duffee, G.R. (2002),
\newblock `Term premia and interest rate forecasts in affine models',
\newblock {\em Journal of Finance\/}~{\bf 57\/}(1), 405--443.

\bibitem[\protect\citeauthoryear{Duffee and Stanton}{Duffee and
  Stanton}{2004}]{ds04}
Duffee, G.R. and R.H.~Stanton (2004),
\newblock `Estimation of dynamic term structure models',
\newblock Manuscript, Haas School of Business.

\bibitem[\protect\citeauthoryear{Duffie and Kan}{Duffie and Kan}{1996}]{dk96}
Duffie, D. and R.~Kan (1996),
\newblock `A yield-factor model of interest rates',
\newblock {\em Mathematical Finance\/}~{\bf 6\/}(4), 379--406.

\bibitem[\protect\citeauthoryear{Fendel}{Fendel}{2005}]{fen05}
Fendel, R. (2005),
\newblock `An affine three-factor model of the {G}erman term structure of
  interest rates with macroeconomic content',
\newblock {\em Applied Financial Economics Letters\/}~{\bf 1\/}(3), 151--156.

\bibitem[\protect\citeauthoryear{Harvey}{Harvey}{1989}]{har89}
Harvey, A.C. (1989),
\newblock {\em Forecasting, structural time series models and the Kalman
  filter}.
\newblock Cambridge University Press.

\bibitem[\protect\citeauthoryear{H\"{o}rdahl, Tristani, and
  Vestin}{H\"{o}rdahl~\emph{et~al.}}{2006}]{htv06}
H\"{o}rdahl, P., O.~Tristani, and D.~Vestin (2006),
\newblock `A joint econometric model of macroeconomic and term-structure
  dynamics',
\newblock {\em Journal of Econometrics\/}~{\bf 131\/}(1-2), 405--444.

\bibitem[\protect\citeauthoryear{Hunt and Kennedy}{Hunt and
  Kennedy}{2000}]{Hunt}
Hunt, P. and J.~Kennedy (2000),
\newblock {\em Financial Derivatives in Theory and Practice}.
\newblock Wiley Series in Probability and Statistics.

\bibitem[\protect\citeauthoryear{Karatzas and Shreve}{Karatzas and
  Shreve}{1991}]{Karshr}
Karatzas, I. and S.~Shreve (1991),
\newblock {\em Brownian Motion and Stochastic Calculus}.
\newblock Springer-Verlag.

\bibitem[\protect\citeauthoryear{Kloeden and Platen}{Kloeden and
  Platen}{1999}]{Kloeden}
Kloeden, P.E. and E.~Platen (1999),
\newblock {\em Numerical Solution of Stochastic Differential Equations}.
\newblock Springer.

\bibitem[\protect\citeauthoryear{Lund}{Lund}{1997}]{lun97}
Lund, J. (1997),
\newblock `Econometric analysis of continuous-time arbitrage-free models of the
  term structure of interest rates',
\newblock Working Paper, Aarhus School of Business.

\bibitem[\protect\citeauthoryear{Musiela and Rutkowski}{Musiela and
  Rutkowski}{1997}]{musrut}
Musiela, M. and M.~Rutkowski (1997),
\newblock {\em Martingale methods in financial modelling}, Volume~36 of {\em
  Applications of Mathematics (New York)},
\newblock Berlin: Springer-Verlag.

\bibitem[\protect\citeauthoryear{Rudebusch and Wu}{Rudebusch and
  Wu}{2007}]{rw07}
Rudebusch, G.D. and T.~Wu (2007),
\newblock `Accounting for a shift in term structure behavior with no-arbitrage
  and macro-finance models',
\newblock {\em Journal of Money, Credit, and Banking\/}, forthcoming.

\bibitem[\protect\citeauthoryear{Spencer}{Spencer}{2004}]{spe04}
Spencer, P. (2004),
\newblock `Affine macroeconomic models of the term structure of interest rates:
  {T}he {US} treasury market 1961-99',
\newblock Discussion Papers in Economics 2004/16, The University of York.

\bibitem[\protect\citeauthoryear{Vasicek}{Vasicek}{1977}]{vas77}
Vasicek, O.A. (1977),
\newblock `An equilibrium characterization of the term structure',
\newblock {\em Journal of Financial Economics\/}, {\bf 5}, 177--188.

\bibitem[\protect\citeauthoryear{Wu}{Wu}{2006}]{wu06}
Wu, T. (2006),
\newblock `Macro factors and the affine term structure of interest rates',
\newblock {\em Journal of Money, Credit, and Banking\/}~{\bf 38\/}(7),
  1847--1875.

\end{thebibliography}

\end{document}